\documentclass[12pt,preprint]{aastex}

\shortauthors{Hao, H. et.~al}

\shorttitle{Variable Intervening Fe~II and Mg~II Absorber}

\begin{document}

\title{Strongly Variable $z=1.48$  Fe~II and Mg~II Absorption in the
Spectra of $z=4.05$ GRB\,060206\altaffilmark{1}}

\author{
H.~Hao\altaffilmark{2}, K.~Z.~Stanek\altaffilmark{3},
A.~Dobrzycki\altaffilmark{4}, T.~Matheson\altaffilmark{5},
M.~C.~Bentz\altaffilmark{3}, J.~Kuraszkiewicz\altaffilmark{2},
P.~M.~Garnavich\altaffilmark{6}, J.~C.~Howk\altaffilmark{6},
M.~L.~Calkins\altaffilmark{7}, G.~Worthey\altaffilmark{8},
M.~Modjaz\altaffilmark{2}, J.~Serven\altaffilmark{8} }

\altaffiltext{1}{Based on data from the F.~L. Whipple Observatory,
  which is operated by the Smithsonian Astrophysical Observatory.}

\altaffiltext{2}{Harvard-Smithsonian Center for Astrophysics, 60
Garden Street, Cambridge, MA 02138; {hhao@cfa.harvard.edu},
{jkuraszkiewicz@cfa.harvard.edu}, {mmodjaz@cfa.harvard.edu}}

\altaffiltext{3}{Department of Astronomy, The Ohio State
University, 140 W. 18th Avenue, Columbus, OH 43210;
{kstanek@astronomy.ohio-state.edu},
{bentz@astronomy.ohio-state.edu}}

\altaffiltext{4}{European Southern Observatory,
Karl-Schwarzschild-Strasse 2, Garching, D-85748, Germany;
{adam.dobrzycki@eso.org}}

\altaffiltext{5}{National Optical Astronomy Observatory, 950 North
Cherry Avenue, Tucson, AZ 85719; {matheson@noao.edu}}

\altaffiltext{6}{Department of Physics, University of Notre Dame,
225 Nieuwland Science Hall, Notre Dame, IN 46556;
{pgarnavi@nd.edu}, {jhowk@nd.edu}}

\altaffiltext{7}{F.~L.~Whipple Observatory, 670 Mt.~Hopkins Road,
P.O.~Box 97, Amado, AZ 85645; {mcalkins@cfa.harvard.edu}}

\altaffiltext{8}{Department of Physics and Astronomy, P.~O.~Box
642814, Washington State University, Pullman, WA 99164;
{gworthey@wsu.edu}, {jdogg@wsu.edu}}

\begin{abstract}

We report on the discovery of strongly variable \ion{Fe}{2} and
\ion{Mg}{2} absorption lines seen at $z=1.48$ in the spectra of
the $z=4.05$ gamma-ray burst (GRB) 060206 obtained between 4.13 to
7.63 hours (observer frame) after the burst.  In particular, the
\ion{Fe}{2} line equivalent width (EW) decayed rapidly from $1.72
\pm 0.25$~\AA\ to $0.28 \pm 0.21$~\AA, only to increase to $0.96
\pm 0.21$~\AA\ in a later spectrum. The \ion{Mg}{2} doublet shows
even more complicated evolution: the weaker line of the doublet
drops from $2.05\pm 0.25$~\AA\ to $0.92\pm 0.32$~\AA, but then
more than doubles to $2.47\pm 0.41$~\AA\ in later data.  The ratio
of the EWs for the \ion{Mg}{2} doublet is also variable, being
closer to $1\!:\!1$ (saturated regime) when the lines are stronger
and becoming closer to $2\!:\!1$ (unsaturated regime) when the
lines are weaker, consistent with expectations based on atomic
physics.  We have investigated and rejected the possibility of any
instrumental or atmospheric effects causing the observed strong
variations. Our discovery of clearly variable intervening
\ion{Fe}{2} and \ion{Mg}{2} lines immediately indicates that the
characteristic size of intervening patches of \ion{Mg}{2}
``clouds'' is comparable to the GRB beam size, i.e., about
$10^{16}\;$cm. We discuss various implications of this discovery,
including the nature of the \ion{Mg}{2} absorbers, the physics of
GRBs, and measurements of chemical abundances from GRB and quasar
absorption lines.

\end{abstract}


\section{Introduction}

The optical spectrum of a gamma-ray burst (GRB) afterglow is
generally a power-law with absorption lines superimposed.  These
absorption lines correspond to intervening material between the
source and the observer. Typically, the highest-redshift lines are
thought to be associated with the host galaxy \citep[e.g., see
discussion in][]{jha01}, and thus provide a measurement of the
redshift of the burst. There might be several lower-redshift
absorption lines caused by intervening systems.  Spectra of
luminous GRB afterglows have opened a new window on
absorption-line studies along lines of sight that are not
associated with quasars \citep[cf.][]{vreeswijk03}.

Recent studies of GRB absorption spectra have produced an
unexpected result. \cite{prochter06} reported statistically
significant evidence for a much higher incidence of strong
\ion{Mg}{2} absorbers along GRB sight lines than that expected
from studies of quasars of comparable redshift. They considered
and rejected three possible scenarios to produce such an effect,
namely dust associated with the \ion{Mg}{2} absorbers, ejection of
\ion{Mg}{2} by the GRB, and strong gravitational lensing by
galaxies associated with the emission.

\cite{frank06} proposed a simple geometric solution to the quandary,
with resulting predictions that are readily testable.  Namely, if the
\ion{Mg}{2} absorbers seen in both GRB and quasar spectra are truly
intervening, then the Prochter et al.\ result could be caused by
different sizes of the GRB and quasar emission regions.  If the
\ion{Mg}{2} absorbers have physical sizes comparable to the emitting
regions of quasars and GRBs, then a background source with a larger
emission region would ``see'' a smaller effective absorption column
density than a background source with a smaller emission region (see
Frank et al.\ for details of their proposed scenario).

One immediate prediction of the Frank et al.\ scenario is that for
strongly dynamic objects like GRB afterglows, the strength and
structure of intervening absorption lines in GRB spectra should vary
in some, if not all, cases. Motivated by this prediction, we have
examined our existing, multi-epoch spectroscopic data of GRB\,060206
and have discovered such a behavior.  In this paper, we provide the
first observational evidence of this prediction. In fact, as far as we
know this is the first report ever of clearly evolving intervening
(i.e., non-intrinsic) ground state absorption lines seen in spectra of
any cosmological object. There is evidence for variation in fine
structure lines in the GRB host environment \citep[cf.][]{vreeswijk06,
zavadsky06}.

GRB\,060206 triggered {\em Swift}-BAT on 2006 Feb 6 at 04:46:53 UT
\citep{morris2006}. A likely afterglow was identified by
\citet{fynbo06}, who also determined that the afterglow was at
high redshift ($z=4.048$).  The evolution of a bright and complex
afterglow was reported by several groups, and \cite{wozniak06},
\cite{stanek06}, and \cite{monfardini06} have published detailed
photometric data for this ``anomalous'' event. The burst also
showed complex behavior in the X-rays as seen with {\em Swift}-XRT
instrument, but the overall X-ray evolution follows that seen in
the optical \citep{stanek06}.

\section{Spectroscopic Observations}

The brightness of the optical transient (OT) associated with
GRB\,060206 allowed us to take spectra for several hours after the
burst.  Spectra of the OT were obtained with the FAST spectrograph
\citep{fabricant98} at the Cassegrain focus of the 1.5m Tillinghast
telescope at the F.~L. Whipple Observatory at 08:25, 09:02, 09:33,
10:05, 10:41, 11:37, 12:09 (all times UT).  The OT was visible on the
telescope acquisition camera and could be identified and placed on the
spectrograph slit for the observations. Each spectrum was taken at or
near the parallactic angle \citep{filippenko82}. The spectra were
reduced in the standard manner with IRAF\footnote{IRAF is distributed
by the National Optical Astronomy Observatory, which is operated by
the Association of Universities for Research in Astronomy, Inc., under
cooperative agreement with the National Science Foundation.} using the
optimal extraction package of \citet{horne86}.  Night-sky emission
lines were removed by fitting a line to two sky regions on either side
of the OT column-by-column; the value determined from the linear
function at the position of the OT extraction was then subtracted.
Wavelength calibration was provided with an HeNeAr lamp taken
immediately after each OT spectrum.  Minor adjustments to the
wavelength solution were made based upon night-sky lines in the OT
frames. There were clouds during some of the exposure, but the
signal-to-noise ratio (S/N) is comparable for all the spectra. The
resolution of each spectrum is $~6.2$~\AA\, as measured from the FWHM
of night-sky emission lines.

We flux-calibrated the spectra with spectrophotometric standard
stars using our own software.  Relative fluxes in our OT spectrum
are accurate to $~4\%$ over the observed wavelength range.
Following the arguments of \cite{wade88} and \cite{matheson00},
telluric correction was accomplished by dividing the GRB spectra
with a normalized spectrum of the flux standard, which was set to
unity for regions unaffected by atmospheric features. For our
spectra, telluric correction was only applied over the regions
6250-6360~\AA\ and 6840-7410~\AA. We scaled the strength of the
absorption by the difference of the airmasses to the $0.6$ power
to correct the saturated B band properly \citep{wade88}. Most of
the correction is $~2\%$ or less, with a $~5\%$ correction in the
6250-6360~\AA\ region and $~25\%$ in the strong region of the B
band.

\begin{figure}
\plotone{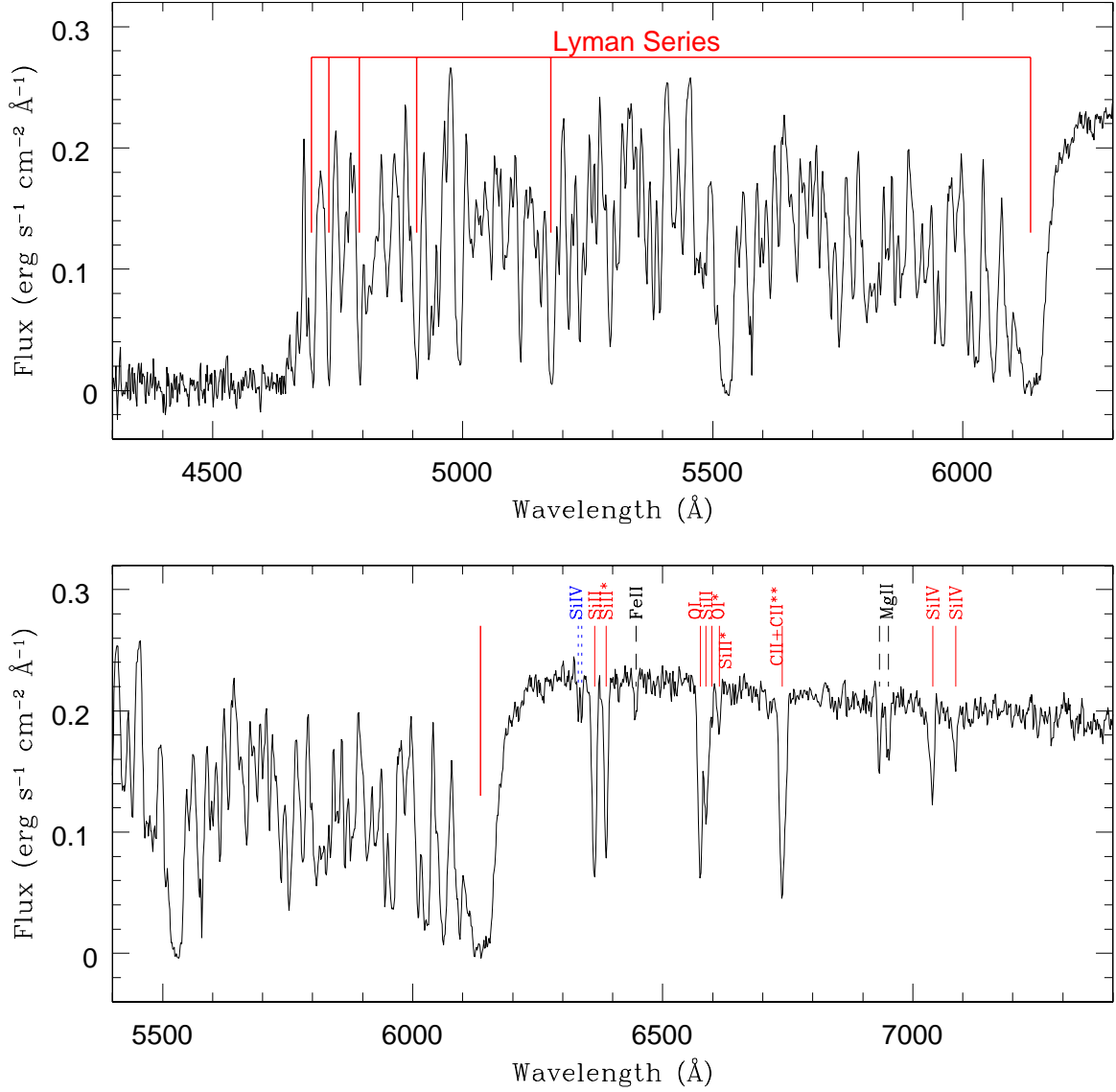} \caption{Spectra of the OT associated with
GRB\,060206. Prominent absorption features associated with three
redshift systems are indicated: $z = 1.48$ (\emph{black long
dashed vertical lines}); $z = 4.05$ (\emph{red solid vertical
lines}); $z=3.54$ (\emph{blue short dashed vertical lines}).
\label{grb-all}}
\end{figure}

In the S/N weighted average spectrum, shown in Figure \ref{grb-all}, a
damped Lyman-alpha absorption system (DLA) is present at
$\lambda=6145$~\AA. A number of strong and weak absorption metal line
systems are superimposed on the continuum, identified as systems at
$z=1.48$ and $z=4.05$ \citep[reported by ][]{fynbo06}. The very strong
DLA system at $\lambda=6145$~\AA\ indicates that the other absorbers
at $z=4.05$ most likely represent the host galaxy of the GRB. In the
intervening metal line system at $z=1.48$, our data show absorption
from the transitions \ion{Fe}{2} $\lambda$2600 and \ion{Mg}{2}
$\lambda\lambda$2796, 2803. These are fairly typical in absorption
line systems \citep[e.g., ][]{steidel92}. Two absorption lines at
$\lambda=6331.3$~\AA\ and $6338.3$~\AA\ represent a \ion{Si}{4} system
at $z=3.54$, with a strong corresponding Ly$\alpha$ absorption line at
$5525.9$~\AA.

\section{Spectroscopic Monitoring}

\begin{figure}
\plotone{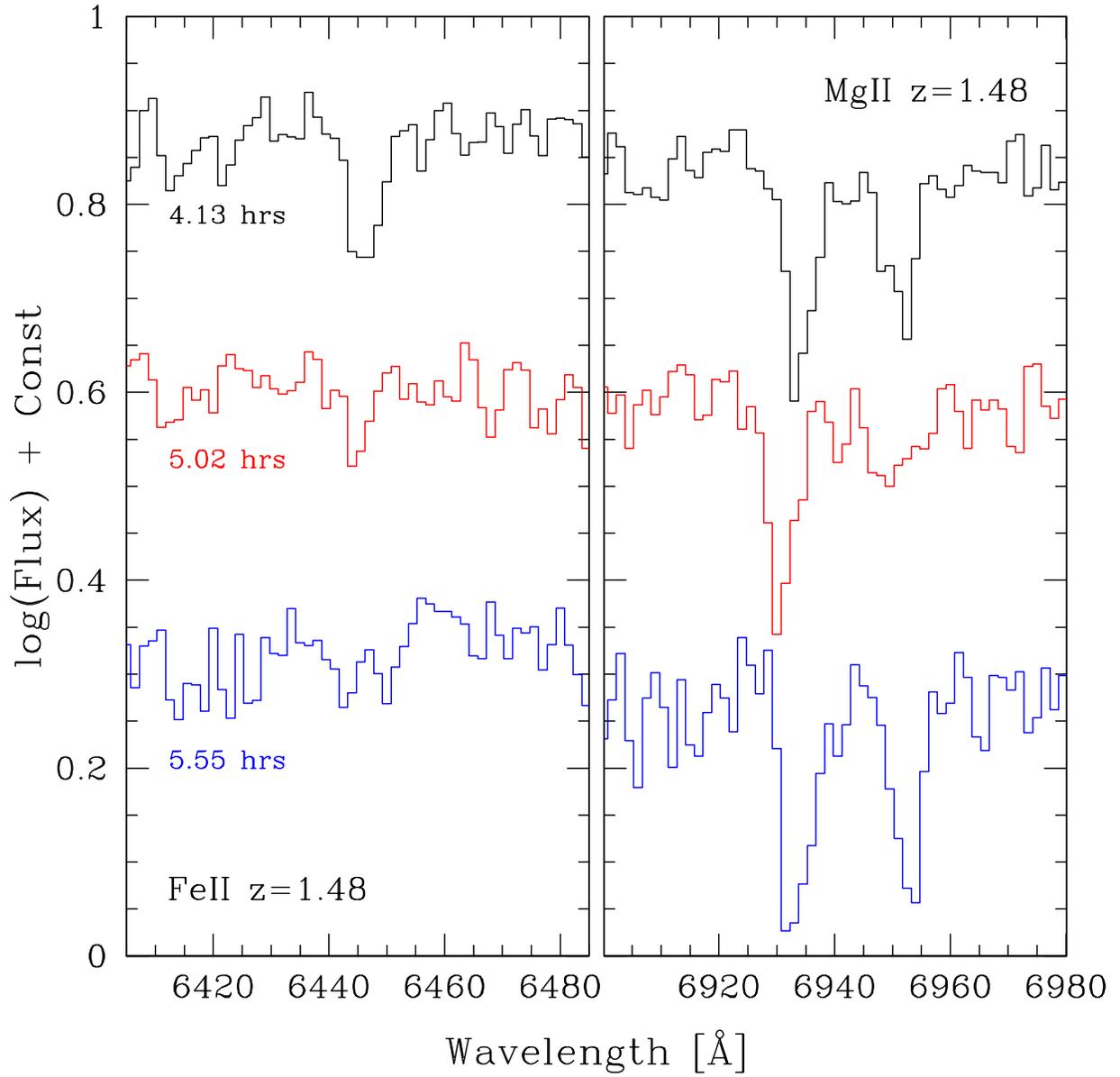}
\caption{Spectra of the GRB\,060206 taken at 4.13
hours, 5.02 hours
         and 5.55 hours after the burst trigger in the $z=1.48$
         \ion{Fe}{2} and \ion{Mg}{2} region. Each spectrum has been
         offset by a constant flux value for clarity.
\label{compare}}
\end{figure}

Motivated by the prediction of Frank et al. (2006) that the
absorption lines along GRB sight lines should vary in some cases,
we examined our time series of GRB~060206 spectra to look for this
effect.  Variation of the intervening $z=1.48$ \ion{Fe}{2} and
\ion{Mg}{2} lines is obvious even upon casual inspection of the
spectra, as shown in Figure~\ref{compare}.  These lines are seen
to vary in strength, profile, and possibly velocity.  We have
checked the original 2-D spectra and no cosmic ray features are
apparent that would cause this variation. In addition, night-sky
emission lines and telluric features were carefully removed with
several variations in technique to ensure that reduction processes
did not introduce the observed variability. In fact, there are no
sky lines or telluric features at all near the strongly variable
$z=1.48$ \ion{Fe}{2}.

\begin{figure}
\plotone{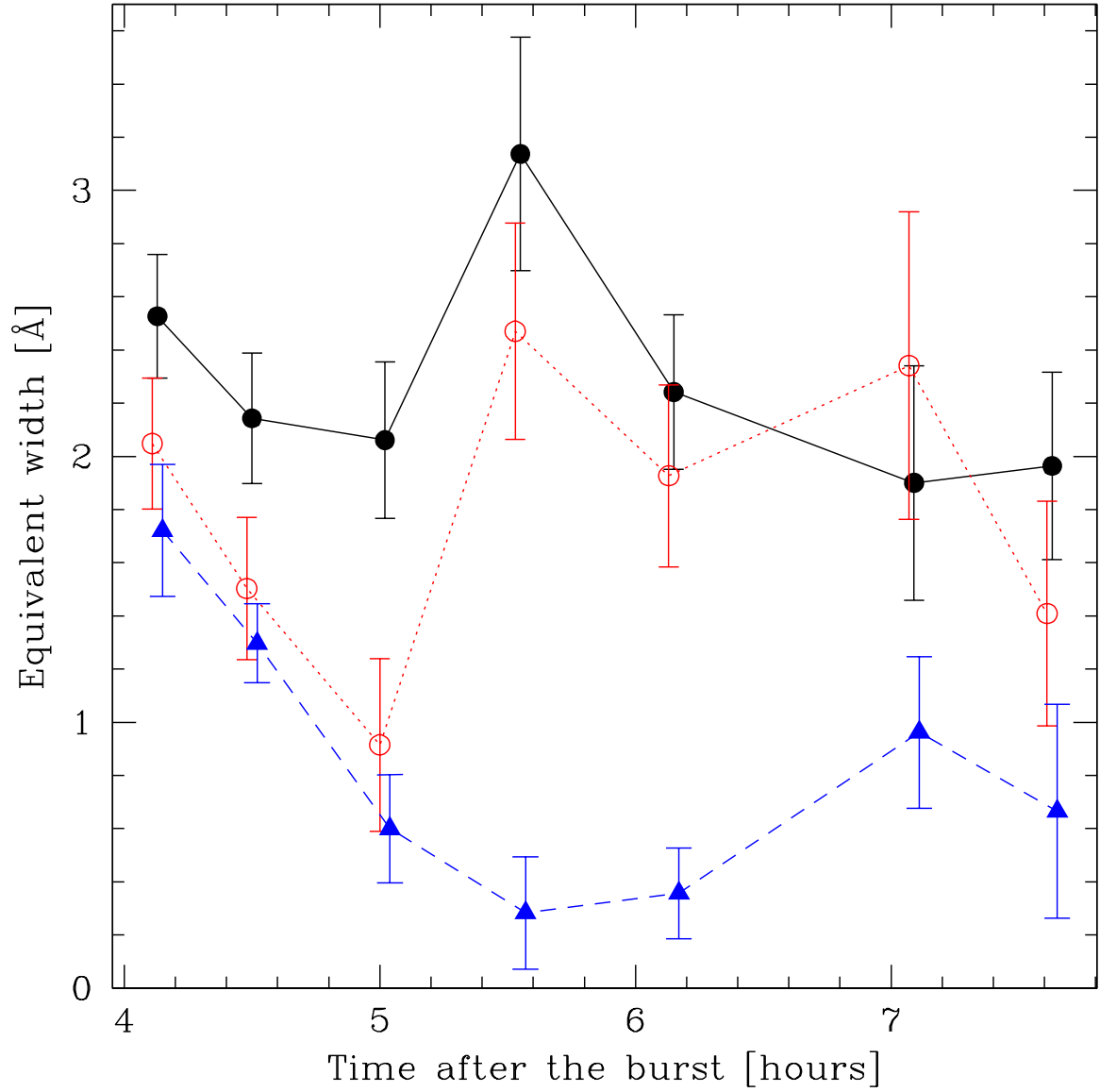}
\caption{Time evolution of the observed EW of the
absorption lines of \ion{Mg}{2} $\lambda$2796 (\emph{filled
circles}) \ion{Mg}{2} $\lambda$2803 (\emph{open circles}), and
\ion{Fe}{2} $\lambda$2600 (\emph{filled squares}). The EW
measurements for the different lines have been slightly offset in
time for clarity.\label{MgIIf}}
\end{figure}

To measure the magnitude of variations in the equivalent widths
(EWs) of the $z=1.48$ absorption lines, we fit each of the
\ion{Mg}{2} lines with multiple Gaussians and the \ion{Fe}{2} line
with a single Gaussian. The error of the EWs are calculated from
the conservative fit uncertainties. The resulting EW measurements
are listed in Table~\ref{MgII} and plotted in Figure~\ref{MgIIf}.
The first three spectra show a gradual decrease in the EW of the
\ion{Mg}{2} $\lambda 2796$ line accompanied by a sharp decline in
the EW of the weaker \ion{Mg}{2} $\lambda 2803$ line.  As a
result, the ratio of the doublet increases from $1.2\!:\!1$ in the
first spectrum to about $2\!:\!1$ in the third.  This is followed
in the fourth spectrum by a sharp rise in the EW of both
components of the doublet and a sharp decrease in the doublet
ratio to $1.3:1$. While the EW of the \ion{Fe}{2} line also
decreases throughout the first three spectra, it seems to
completely disappear in the fourth spectrum. The two small
features in this spectrum at the location of the \ion{Fe}{2}
feature are consistent with noise (see Figure~\ref{compare}).  The
\ion{Fe}{2} absorption line gradually returns throughout the last
three spectra, but it never regains its original strength.  The
only other spectral feature to show possible variability is the
\ion{Si}{4} $\lambda 1393.76$ absorption line at the presumed
redshift of the host galaxy. The variability in this line does not
follow the behavior of either the \ion{Mg}{2} or \ion{Fe}{2}
systems at $z = 1.48$. We are still investigating the nature and
significance of variability in this line.

Variation in the EW of an absorption line is a direct consequence
of variation in the column density of the absorber as seen by the
background source.  The \ion{Fe}{2} and \ion{Mg}{2} systems are
not necessarily expected to be distributed homogeneously
throughout an absorbing ``cloud.''  As the beam size of the GRB
increases due to its physical expansion, the beam could include
various dense or dilute regions of the ``cloud,'' leading to
different behaviors in the measured EWs of the absorbers.
Additionally, the \ion{Fe}{2} and \ion{Mg}{2} may not be
co-spatial, so their behaviors may not be directly comparable as
seen here.

To estimate the statistical significance of these observed
variations, we apply $\chi^2$ statistics to the ``null
hypothesis'' that the \ion{Fe}{2} and \ion{Mg}{2} lines are not
variable.  Fitting the multiple measurements of the \ion{Fe}{2} EW
with a constant value gives $\chi^2=38.8$ with six degrees of
freedom.  Such a value of $\chi^2$ results in a probability of
$p=0.0000794\%$ ($\sim 5 \sigma$ Gaussian significance) that the
\ion{Fe}{2} line is not varying.  A similar analysis for the
\ion{Mg}{2} $\lambda 2803$ line results in $\chi^2 = 14.0$, which
gives $p = 2.93\%$ ($2.2 \sigma$ Gaussian significance) for lack
of variation in the line EW.  Finally, for the \ion{Mg}{2}
$\lambda 2796$ line, we find $\chi^2=7.4$, resulting in $p=28.9\%$
for the assumption that the EW does not vary. The combined data clearly show statistically
significant evidence for variability.  Note that while we only
test here the significance of variations in the line EWs, it is
also immediately apparent from the spectra that the line profiles
are varying as well (see Figure~\ref{compare}).  Each of these
lines is likely composed of several components that are blended in
our spectra, and it is variations in individual components that
are causing the observed variations in the total EWs and line
profiles.

To further investigate the variability of the lines, we use the
spectral subtraction algorithm developed by Hartman \& Stanek (2007,
in preparation), which uses the entire spectrum in the wavelength range
$6250\AA \sim 7200\AA$ to obtain the convolution kernel for matching individual
observed spectra to the reference spectrum. We choose the first
observed spectrum as the reference, then calculate the difference
between the matched spectra and the observed spectra for the rest of
the observed spectra, divided by the variance spectra we can calculate
the significance of the difference. The \ion{Fe}{2} shows a
$4.9\sigma$ variation in the line region $(6439\AA\sim6450\AA)$ and
the \ion{Mg}{2} $\lambda 2796$ line shows a $2.5\sigma$ variation in
the line region $(6926\AA\sim6942\AA)$.

\section{Discussion}

Variable intervening \ion{Fe}{2} and \ion{Mg}{2} was predicted by
\cite{frank06} as a direct consequence of their geometric solution to
the \ion{Mg}{2} absorption over-abundance problem discovered by
\cite{prochter06}. Our result immediately provides constraints on the
physical size of the \ion{Mg}{2} absorbers that are several orders of
magnitude stronger than those determined from studies of lensed
quasars --- $\sim 0.01$~pc \citep{frank06} vs. $< 200-300$~pc
\citep{rauch02, petitjean00}. Observations of lensed quasars such as
those proposed by \cite{dong06} may provide additional constraints on
the physical sizes of the absorbers.  Significant spectral variability
on the timescales of months to years induced by microlensing can be
used to probe the $10^{14}-10^{16}\;$cm cloudlets along the sightlines
of lensed quasars. Properties of the cloudlets (physical size,
sub-structure, etc.) can be studied statistically if such spectral
variations are seen.

We can also investigate the average physical density of the
absorbers. Assuming the ``cloud'' is optically thin, we can obtain
a lower limit on the \ion{Mg}{2} column density $N_{\rm MgII}=
1.7\times10^{13}\;{\rm{cm}}^{-2}$. To evaluate this result, we
have used our third spectrum taken $\sim 5\;$hours after the
burst, where the \ion{Mg}{2} line ratio is close to $2\!:\!1$,
indicating the lines are not strongly saturated (see
Table~\ref{MgII}). The beam size of the GRB during that epoch can
be estimated \citep{waxman97,loeb98} via
\begin{equation}
R_s(t) = 4.1\times10^{15}\left(\frac{E_{52}}{n_1}\right)^{\frac 1
8}
         (1+z_s)^{-5/8}(t/hour)^{5/8}\; {\rm{cm}}
\end{equation}
where $z_s$ is the source redshift, $E_{52}$ is the
``isotropic-equivalent'' of the energy release in units of
$10^{52}~{\rm erg~s^{-1}}$, and $n_1$ is the ambient gas density in
units of ~$1~{\rm cm^{-3}}$. For GRB~060206, $E_{52}=5.8$
\citep{palmer06}, and assuming $n_1 = 1\;{\rm{cm}}^{-3}$, then
$R_s=5.1\times 10^{15}\;$cm for our third spectrum. With the physical
size of the individual \ion{Mg}{2} ``clumps'' constrained to be
comparable to the GRB beam size (i.e., $\sim 10^{16}\;$cm), one can
immediately derive the average physical density of \ion{Mg}{2} atoms
to be about $0.0015~\rm{cm}^{-3}$. This is the first time that such a
strong constraint has been obtained. Given its large \ion{Mg}{2}
equivalent width, the z=1.48 absorber is likely a Lyman limit system
with $\log{N(HI)} > 18$, although there is a $\sim 20-30$\%
probability that it is a damped Lyman-alpha system (Rao et
al. 2006). Then the physical density of the hydrogen of this cloud is
$>200~\rm{cm}^{-3}$. The estimate of physical density of \ion{Mg}{2}
absorbers is very important for understanding the cloud origins and
properties, and more generally the metal lines and metal enrichment
(see discussion in Schaye et al 2007).

Another consequence of this discovery relates to GRBs themselves.  In
the case of GRB\,060206, we see the \ion{Mg}{2} EW initially
decreasing, but then it shows a dramatic increase.  This behavior is
either a result of the GRB beam becoming much smaller in a short time
(possibly due to an episode of energy injection) or part of the
afterglow ring suddenly brightening. Our discovery points the way to a
new method for studying GRB phenomena by monitoring afterglows with
high-resolution optical spectroscopy.

Additional implications relate to chemical abundances for intervening
systems as measured using absorption lines. In our first spectrum, the
ratio of EWs for \ion{Fe}{2} compared to \ion{Mg}{2} is about 0.7.  In
our fourth spectrum, where the \ion{Fe}{2} is only marginally present,
and \ion{Mg}{2} is very strong, that ratio is less than 0.1. Estimates
of relative abundances for these two elements would be dramatically
different when analyzing these two low-resolution spectra. If the
system is at the lower end of this column density range$\log{N(HI)} >
18$, some of the difference in \ion{Mg}{2}/\ion{Fe}{2} ratio could be
the result of spatially-varying ionization levels.  However, it may be
that there are intrinsic variations in the relative metal abundances
(although such variations are not seen in damped Lyman-alpha systems:
Prochaska 2003, Rodriguez et al 2006).

The observations presented in this paper open up the exciting
possibility of studying time-variability of intervening absorption
lines in the spectra of cosmological objects. Strongly and rapidly
variable objects, such as GRBs and blazars, are obvious targets for
further studies, but quasars (including both lensed and non-lensed)
should also be investigated. Variable narrow absorption lines have
been observed in spectra of quasars (e.g., Hamann et al. 1997), but
they have been interpreted as originating in gas physically associated
with the quasar. In fact, variability of narrow absorption lines is
now used as one of the criteria to select intrinsic absorbers
associated with quasars.  Results presented in this paper suggest that
using this criterion should be re-examined---quasars are optically
variable sources, so size of their continuum emission region is bound
to vary, just like it does for GRBs. As already discussed by Frank et
al. (2006), quasars with smaller continuum emitting regions (i.e.,
intrinsically fainter ones), should have higher incidence of the
strong \ion{Mg}{2} absorbers.

Note that throughout this paper we considered the $z=1.48$ variable
absorption system not to be physically associated with the $z=4.05$
GRB, i.e., to be truly intervening due to cosmological expansion. For
these two objects to be physically associated would require
\ion{Fe}{2} and \ion{Mg}{2} absorbing gas to have ejection velocity
$v\sim 0.6\;$c, while maintaining relatively low temperature in order
not to be completely ionized.  We consider such scenario most
unlikely. Detection of a galaxy associated with this particular
$z=1.48$ variable absorption system would settle this issue beyond any
reasonable doubt.  As discussed in Prochter et al. (2006), such
foreground galaxies were already identified for a number of strong,
intervening \ion{Mg}{2} absorbers seen in GRB spectra.

\acknowledgments

We thank J.~Hartman, P.~Martini, S.~Gaudi, S.~Mathur, R.~Pogge, S.~Frank,
X.~Dai, J.~Prieto, D.~An, M.~Elvis, R.~Kirshner, and R.~Narayan
for helpful comments and discussions.  M.~C.~Bentz is supported by
a National Science Foundation Graduate Fellowship.

\clearpage

\begin{deluxetable}{llccc}
\tablewidth{0pt} \tablenum{1} \tablecaption{EVOLUTION OF
EQUIVALENT WIDTH OF THE $z=1.48$ ABSORPTION LINES \label{MgII}}
\tablehead{\colhead{$t_{obs}$\tablenotemark{a}} &
\colhead{Age\tablenotemark{b}} &
\colhead{EW\tablenotemark{c}$_{\rm{Mg~II}~2796}$} &
\colhead{EW$_{\rm{Mg~II}~2803}$} &
\colhead{EW$_{\rm{Fe~II}~2600}$} } \startdata
 2.86384 & 4.13 & 2.53 $\pm$ 0.23 & 2.05 $\pm$ 0.25 & 1.72 $\pm$ 0.25 \\
 2.88946 & 4.50 & 2.14 $\pm$ 0.25 & 1.50 $\pm$ 0.27 & 1.30 $\pm$ 0.15 \\
 2.91110 & 5.02 & 2.06 $\pm$ 0.30 & 0.92 $\pm$ 0.33 & 0.60 $\pm$ 0.20 \\
 2.93323 & 5.55 & 3.14 $\pm$ 0.44 & 2.47 $\pm$ 0.41 & 0.28 $\pm$ 0.21 \\
 2.95857 & 6.15 & 2.24 $\pm$ 0.29 & 1.93 $\pm$ 0.34 & 0.36 $\pm$ 0.17 \\
 2.99745 & 7.09 & 1.90 $\pm$ 0.44 & 2.34 $\pm$ 0.58 & 0.96 $\pm$ 0.29 \\
 3.01990 & 7.63 & 1.96 $\pm$ 0.35 & 1.41 $\pm$ 0.42 & 0.67 $\pm$ 0.40
\enddata
\tablenotetext{a}{Heliocentric Julian Date at middle of exposure
minus 2,453,770.}

\tablenotetext{b}{Age of the middle of exposure in hours from the
detection of the burst at 2006 Feb 06 04:46:53 UT
\citep{morris2006}.}

\tablenotetext{c}{Observer-frame equivalent width.}.

\end{deluxetable}

\clearpage

\clearpage

\clearpage

\end{document}